\documentclass[epsfig,useAMS,useAMSmath, usenatbib]{mn2e}

\usepackage{url}

\usepackage {graphics}
\usepackage{graphicx}
\usepackage {layout}

\newcommand{\Msun}{M_\odot}

\newcommand{\beq}{\begin{equation}}
\newcommand{\eeq}{\end{equation}}

\newcommand\aj{AJ}
\newcommand\apj{ApJ}

\newcommand\apjl{ApJ}
\newcommand\apjs{ApJS}
\newcommand\mnras{MNRAS}

\newcommand\nat{Nature}

\title[Massive galaxies at high redshift]{The fate of high redshift
  massive compact galaxies in dense environments}

\author[Kaufmann et al.]  {Tobias Kaufmann$^{1}$ \thanks{E-mail:
    tobias.kaufmann@phys.ethz.ch}, Lucio Mayer$^2$, Marcella
  Carollo$^{1}$, and Robert Feldmann$^{3,4}$ \\$^1$Institute of
  Astronomy, ETH Zurich, CH-8093 Zurich, Switzerland
  \\$^2$Institute for Theoretical Physics, University of Zurich,
  CH-8057 Zurich, Switzerland \\$^3$Center for Particle
  Astrophysics, Fermi National Accelerator Laboratory, Batavia, IL
  60510, USA \\$^4$Kavli Institute for Cosmological Physics, The
  University of Chicago, Chicago, IL 60637 USA}

\begin{document}

\pagerange{\pageref{firstpage}--\pageref{lastpage}} \pubyear{} 

\maketitle

\begin{abstract}
  Massive compact galaxies seem to be more common at high redshift
  than in the local universe, especially in denser environments. To
  investigate the fate of such massive galaxies identified at $z\sim
  2$ we analyse the evolution of their properties in three
  cosmological hydrodynamical simulations that form virialised galaxy
  groups of mass $\sim 10^{13} M_{\odot}$ hosting a central massive
  elliptical/S0 galaxy by redshift zero. We find that at redshift
  $\sim 2$ the population of galaxies with $M_*>2 \times 10^{10}\Msun$
  is diverse in terms of mass, velocity dispersion, star formation and
  effective radius, containing both very compact and relatively
  extended objects. In each simulation all the compact satellite
  galaxies have merged into the central galaxy by redshift 0 (with the
  exception of one simulation where one of such satellite galaxy
  survives).  Satellites of similar mass at $z=0$ are all less compact
  than their high redshift counterparts. They form later than the
  galaxies in the $z=2$ sample and enter the group potential at $z <
  1$, when dynamical friction times are longer than the Hubble
  time. Also, by $z=0$ the central galaxies have increased
  substantially their characteristic radius via a combination of in
  situ star formation and mergers. Hence in a group environment
  descendants of compact galaxies either evolve towards larger sizes
  or they disappear before the present time as a result of the
  environment in which they evolve. Since the group-sized halos that
  we consider are representative of dense environments in the
  $\Lambda$CDM cosmology, we conclude that the majority of high
  redshift compact massive galaxies do not survive until today as a
  result of the environment.

\end{abstract}


\section{Introduction}

High redshift massive galaxies are observed to have a wide range of
properties. Van Dokkum et al. (2009) report on a massive compact
galaxy at redshift $z=2.186$ with velocity dispersion $\sim 500$ km
s$^{-1}$, stellar mass of $\sim 2\times 10^{11}\Msun$ and an effective
radius of $\sim0.8$ kpc and van de Sande et al. (2011) present a
compact galaxy with dynamical mass of $\sim 1.7\times 10^{11}\Msun$
and velocity dispersion of $\sim 300$ km s$^{-1}$ at redshift $1.8$.
In a complete sample of luminous early type galaxies in the Hubble
Ultra Deep Field of Daddi et al. (2005) roughly half of the galaxies
in the sample have effective radii of $< 1$ kpc (see also Szomoru et
al. 2010) while Mancini et al. (2010) derive for a sample of 12 ultra
massive early-type galaxies at $ 1.4 < z < 1.7$ effective radii
comparable to those of local ellipticals. Also Onodera et al. (2010)
report the detection of a massive galaxy at $z=1.82$ with properties
fully consistent with those of today's giant ellipticals. In the local
universe the massive, compact objects seem not exist anymore (see
e.g. the SDSS sample, York et al. 2000, as presented in van de Sande
et al. 2011) and at fixed stellar mass early-type galaxies were
generally more compact and denser at earlier times (Cappellari et
al. 2009, van de Sande et al. 2011). However, consensus in this debate
has not yet been reached, Szomoru, Franx \& van Dokkum (2011) found
that the number density for passively evolving massive compact
galaxies declines with time whereas new work (Carollo et al in prep.)
shows that this number density constant stays versus redshift.

Mechanisms to grow a compact elliptical galaxy in size have been
investigated in the literature, such as accretion of stars from minor
and major mergers as well as redistribution of angular momentum of the
in situ stellar component (e.g. Khochfar \& Silk 2006, Naab et
al. 2007, 2009; Oser et al. 2010; Hopkins et al. 2010; Bezanson et
al. 2009; Feldmann et al. 2010, F10 hereafter). Such mechanisms can in
principle turn an ultra-compact high-z galaxy into an early-type
galaxy with a much larger effective radius and lower density,
comparable to that of present-day ellipticals.  Recently, Oser et
al. (2011) have used 40 cosmological re-simulations of massive,
individual ('field') galaxies to show that the simulated galaxies
having $(M_*> 10^{11}\Msun)$ at $z=2$ are compact with high velocity
dispersion. Those galaxies then grow in size until $z=0$ mostly due to
minor mergers\footnote{The importance of merging for size growth has
  also been pointed out by recent observational work (e.g., Bluck et
  al. 2011, Whitaker et al. 2011, Newman et al. 2011,
  M{\'a}rmol-Queralt{\'o} et al. 2012).}  to become
more consistent with the local (SDSS) sizes. However, it is unclear
whether galaxy formation simulations in the context of the
$\Lambda$CDM model can reproduce not only the existence of extremely
compact galaxies at high redshift but also the fact that a large
spread in the properties of massive galaxies already exists at high
redshift based on the latest observations. In addition, little is
known about the connection between early-type galaxies existing at low
and high redshift besides the difference in the typical densities and
effective radii.  Simulations have the potential to shed light on this
issue, as they have done already in the case of disc galaxies (e. g.
Brooks et al. 2010). F10 have studied the evolution of massive
early-type galaxies at the centre of virial groups with mass $\sim
10^{13} M_{\odot}$ in cosmological simulations analysing their
structural evolution from $z\sim1.5$ to $0$. In a complementary study,
Feldmann, Carollo \& Mayer (2011) (FCM11 hereafter) have investigated
the environmentally-driven evolution of the non-central group members
in one of these groups focussing on the galaxy population present at
$z=0.1$. Here, on the other hand, we discuss the properties of
galaxies identified to be massive $(M_*>2\times10^{10} \Msun)$
\emph{at $z\geq 2$}, comparing them with observations of massive
galaxies at $z=2$ and establishing the evolutionary connection with
the final member galaxies of the groups at $z=0$. We note that each of
these three simulations of group haloes shows several ($\sim 5$)
massive galaxies already at redshift 2.

We show that at redshift $\sim 2$ the population of massive galaxies
is diverse in terms of mass, velocity dispersion and effective radius
(although generally more compact than the local counterparts), in
agreement with the picture emerging from the observations. Despite
their variety at high redshift in all the three simulations the main
progenitors evolve into fairly typical massive early-type galaxies at
redshift zero, with similar stellar masses, sizes and velocity
dispersions.  We discuss the implications of the latter result in the
general galaxy formation picture. We find that all (but one) of the
massive galaxies selected at redshift $2$ merge to form the most
massive central galaxy at redshift zero in each simulation. Today's
massive satellite galaxy population did not exist already at high z.
The massive satellite galaxies selected at redshift 0 acquire most of
their stellar mass much later than $z=2$ and are found to be less
compact than the high redshift sample.

In Section 2, we present our initial conditions, numerical techniques
and the methodology for the analysis. In Section 3, we discuss the
evolution of the properties of the massive galaxies selected at
redshift two and zero. Section 4 discusses the role of formation time
and the influence of missing physics and numerical resolution.  We
conclude and summarise in Section 5.

\section{Simulations}

We analysed a set of three cosmological smoothed particles
hydrodynamics (SPH) simulations at the galaxy group scale originally
presented and described in F10.  The groups called G1, G2 and G3 have
similar virial masses at redshift 0 ($\sim 10^{13}\Msun$) but
different merger histories and environments.

Those galaxy groups were selected from a DM-only simulation (Hahn et
al. 2007) based on their halo masses. The re-simulation of those
patches were performed in the WMAP3 cosmology (Spergel et al. 2007)
using several layers of resolution enclosing each galaxy group with
gas particles added to the highest resolution regions. The initial
power spectrum has been calculated using \textsc{linger} (Bertschinger
1995) and the refinements were generated using \textsc{grafic-2}
(Bertschinger 2001). All groups were evolved to redshift $0$ at
standard resolution where the dark matter has been sampled with
particles of mass $3.7\times 10^7 \Msun \,h^{-1}$ (where $h=0.73$) and
the gas with particles having initial mass of $7.9\times 10^6 \Msun
\,h^{-1}$. The gravitational (spline) softening length used was $0.73$
and $0.44$ kpc $h^{-1}$ for the dark and baryonic particles,
respectively. Additionally, a high-resolution version of G2 was
evolved down to z $= 0$ using $\sim$ eight times better mass and
$\sim$ two times better force resolution. In this paper we are
reporting the results from the high-resolution version of G2 and from
the standard resolution versions of G1 and G3. We additionally use the
standard resolution run of G2 to analyse the influence of numerical
resolution.

The simulations were performed using the parallel TreeSPH code
\textsc{Gasoline} (Wadsley et al. 2004). The code includes radiative
cooling for a primordial mixture of helium and (atomic) hydrogen.
Because of the lack of molecular cooling and metals, the efficiency of
our cooling functions drops rapidly below $10^4$ K. Star formation and
feedback is modelled as in Stinson et al. (2006); stars spawn from
cold, Jeans unstable gas particles in regions of converging
flows. Once a gas particle is eligible for spawning stars, it does so
based on a probability distribution function with a star formation
rate parameter $c^*=0.05$ that has been tuned to match the Kennicutt
(1998) Schmidt Law. Each star particle is treated as a single stellar
population with Scalo IMF (Miller \& Scalo 1979). Feedback from
supernovae Type Ia and II is included in the simulation. The latter
are modelled using the blastwave scenario from McKee and Ostriker
(1977), which involves shutting-off the cooling for gas encompassed by
the blast-wave over a duration comparable to the Sedov plus snowplaugh
phases (10-20 Myr). Such model for star formation and feedback has
proven to be successful in simulating the formation of realistic disc
galaxies at both low and high mass scales (Mayer, Governato \&
Kaufmann 2008; Governato et al. 2010; Guedes et al. 2011). At
sufficiently high resolution it is even possible to form disc galaxies
and ellipticals/S0s in the same simulation (FCM11).

In all the simulations a population of relatively massive main
progenitors, which are star forming and host reservoirs of cold gas,
evolve to massive, gas-poor early-type systems supported by stellar
velocity dispersion. By redshift zero those central galaxies are
resembling either elliptical or S0 galaxies.

\subsection{Methodology of the analysis}

We select all galaxies at redshift $2$ with stellar masses $>2 \times
10^{10}\Msun$ using a friends of friends (fof) algorithm with a
linking length of $0.3$ kpc. The stellar masses of those galaxies are
calculated by adding up all stellar particles in a sphere of radius
$10$ (physical) kpc around the centre of the stellar particles. At
redshift $0$ all quantities are calculated within a radius of $20$ kpc
around the centre. While the choice of this radius is arbitrary to
some extent the changes in stellar masses stay small when the radius
is varied within a factor of $\sim2$ (F10). This has been quantified
in FCM11, who showed that a radius of $20$
kpc encloses $\sim 98\%$ of the stellar mass of the galaxies at
$z=0.1$. We correct for star formation in the unresolved centres of the
galaxies following the {\it minimal star formation correction
  approach} described in detail in the appendix of F10. Masses and
effective radii derived using the correction are shown as error bars
in the Fig. \ref{fig2a}.

We define the effective radius as the radius which includes half of
the stellar mass within $20$ (physical) kpc at $z=0$ ($10$ kpc for
$z>0$) around the centre of the respective galaxy. The stellar
velocity dispersion has been calculated along a randomly chosen line
of sight (LOS) through the galaxy and as well along two additional LOS
orthogonal to the others. The quoted velocity dispersion is averaged
over all the results from the different LOS and errors come from the
difference of the average and the minimal (maximal) value of
dispersions, respectively.

Additionally, in the high-resolution version of G2 we select a
population of massive galaxies (referred to as 'satellite' galaxies)
at redshift 0 with stellar masses $> 2 \times 10^{10}\Msun$, i.e. with
masses equivalent to those of the $z=2$ sample.  The central galaxy
has been excluded from the $z=0$ sample. These five satellites are
then traced back to $z=2$. All the quantities are calculated within a
radius of $10$ kpc around the centre at all redshifts given that those
galaxies do not extend beyond that significantly. We have adapted the
minimal star formation correction approach of F10 for the satellites:
The average minimal amount of star formation of $0.33\, \Msun$
Gyr$^{-1}$ in the satellites within the inner softening length (which
is potentially artificial) has been removed from the inner region, as
described in the appendix of F10.  Again, values derived using the
correction are shown as error bars in the respective figure.

\section{The evolution of massive galaxies}

\subsection{The nature of massive galaxies at redshift 2}

\begin{figure}
\includegraphics[scale=0.95]{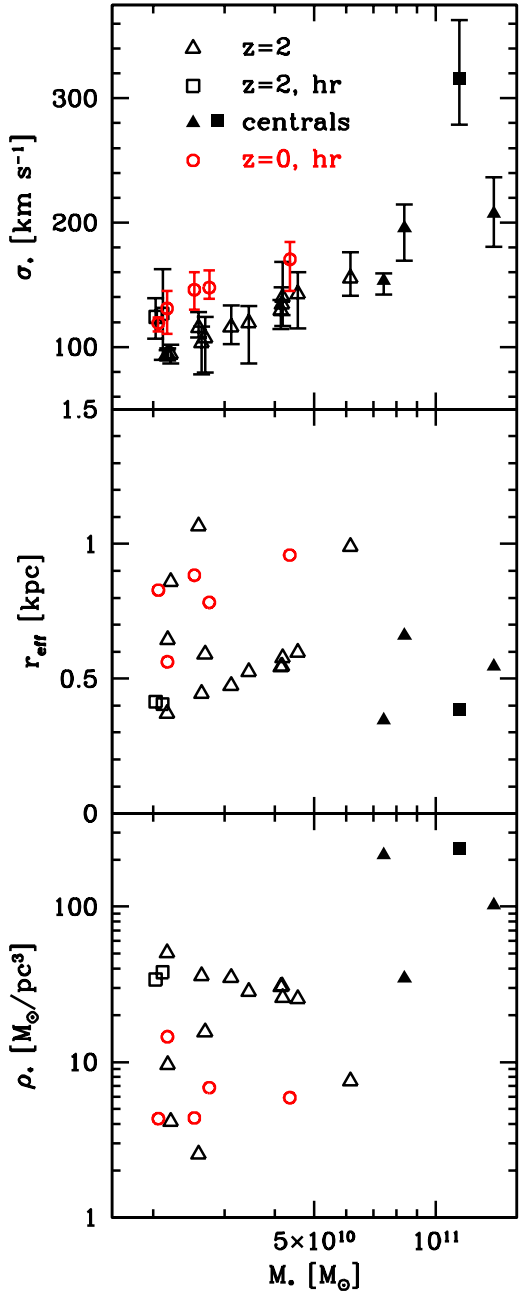}
\caption{The massive galaxy population at redshift 2 (black) and the
  massive satellites at redshift 0 (red). From top to bottom plot: the
  stellar velocity dispersion measured within the effective radius
  versus stellar mass, effective radius versus stellar mass and the
  effective density (stellar density within the effective radius)
  versus stellar mass are shown. The measurement for the galaxies of
  the three different simulations are plotted using triangles
  (standard resolution) or squares (high resolution). Filled symbols
  are used for the galaxies evolving into the most massive, central
  object until redshift 0 in each of the simulations (the 'main
  progenitors'). Circles are used for the population of massive
  (satellite) galaxies selected at redshift 0 in the high resolution
  simulation.  The progenitors of today's ellipticals show a big
  spread in mass and velocity dispersion and are generally more
  compact than the massive z=0 (satellite) galaxies.\label{fig1}}
\end{figure}

The 16 massive galaxies identified at redshift 2 (6 galaxies in G1, 3
in G2, and 7 in G3) form loose associations with maximal distances
between the most massive and any other selected galaxy of up to $\sim
1100$ physical kpc\footnote{In this maximal case, G1, all but one of
  the selected galaxies were distributed along one of the main
  filaments.} Only one of those galaxies lies within the virial radius
of the main progenitor of one of the later central galaxies (in group
G1).  Fig. \ref{fig1} shows the spread in properties (stellar mass and
stellar velocity dispersion) of the selected galaxies. The galaxies
have masses between $\sim 2 \times 10^{10}\Msun$ and $1.15 \times
10^{11}\Msun$ within the central $10$ (physical) kpc and velocity
dispersions from $\sim 95$ to $\sim 320$ km/s within the effective
radius with rather big errors (i.e., results are dependent on the line
of sight). Note that also the degree of rotational support of those
galaxies vary: one of the main progenitors showed a $v/\sigma < 0.5$,
whereas the other main progenitors are having values $>0.8$ and many
of the (especially the lower mass) galaxies reach values of $v/\sigma
\sim 2$ (Fig. \ref{fig2b}) and show a disc-type morphology. The gas
content $m_{gas}/m_{baryons}$ within the 10 kpc sphere lies between
$\sim 14 \%$ and$\sim 27 \%$ for the later satellite galaxies, for the
later central galaxies the gas fractions are $\sim 14 \%$ for G1 and
G3 and go down to $\sim 2 \%$ in the case of early forming (see
section 4.1) central object of G2.  The most massive progenitor of the
central object at $z=0$ of group G2 already reaches $1.15 \times
10^{11}\Msun$ at redshift 2. This, combined with its high velocity
dispersion and its small effective radius (see Fig. \ref{fig2a}),
makes this object similar to the observations by van Dokkum et
al. (2009), although the galaxy formed in G2 shows smaller values of
mass, velocity dispersion and effective radius. Note, that at redshift
2 none of the G2 satellites selected at redshift 0 (see section
\ref{sat_0} has a stellar mass $>2 \times 10^{10}\Msun$ (as would be
needed for selection at $z=2$). In fact, all of those satellites have
stellar masses $<2 \times 10^{10}\Msun$ already at $z=1.5$, see
Fig. \ref{fig2a}.
 
\subsection{The time evolution of massive galaxies selected at redshift 2}

\begin{figure*}
\includegraphics[scale=1.0]{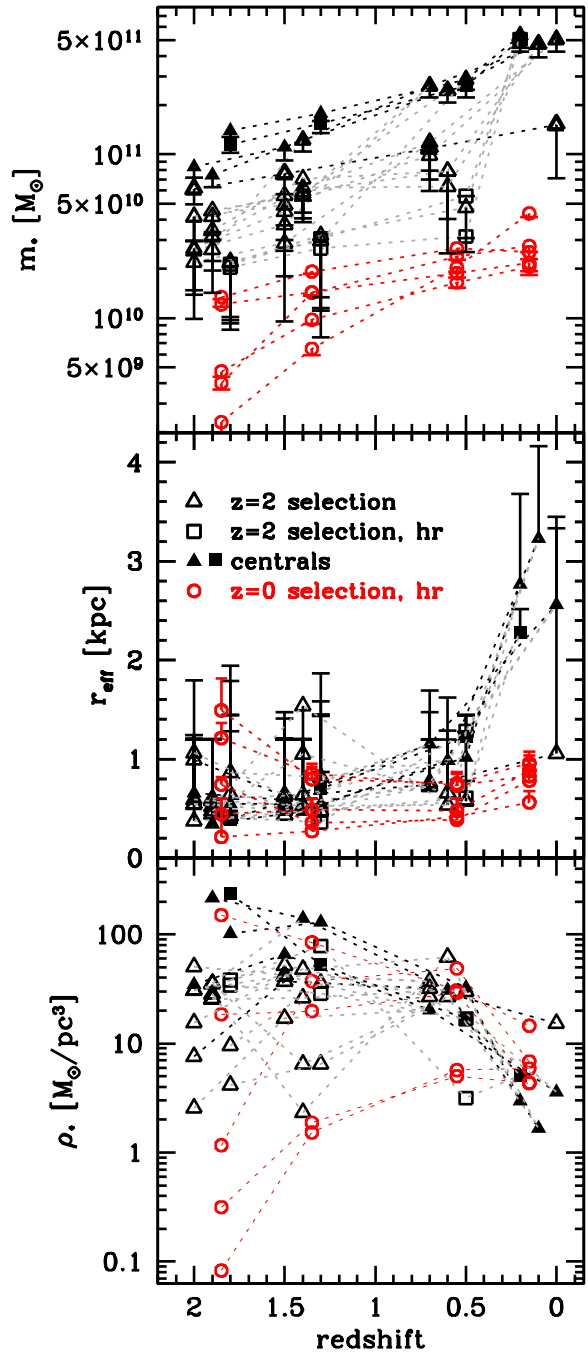}
\caption{The time evolution of stellar mass, effective radius and
  effective density of the massive galaxy population selected at
  redshift 2 (black) and at redshift 0 (red) is shown.  From top to
  bottom plot: Stellar mass, effective stellar radius and effective
  density (stellar density within the effective radius) measured
  redshift 2, 1.5, 0.7 and 0 are shown. Measurements were taken at the
  redshifts indicated above but are plotted shifted slightly along the
  x-axis for better visibility. The measurement for the galaxies of
  the three different simulations are plotted using triangles
  (standard resolution) or squares (high resolution). Filled symbols
  are used for the galaxies evolving into the most massive, central
  object until redshift 0 in each of the simulations (the 'main
  progenitors'). Circles are used for the population of massive
  (satellite) galaxies selected at redshift 0 in the high resolution
  simulation. Lines are connecting the same object
  over time (until it is merged to an another galaxy). Galaxies
  surviving until redshift 0 are indicated by thick lines. Note the
  very similar masses of the main galaxies at redshift 0.  \label{fig2a}}
\end{figure*}

In all the simulations all of the massive galaxies selected at
redshift 2 merge subsequently to one massive galaxy at redshift 0 in
the respective simulation, with only one exception of one additional
galaxy surviving. This is illustrated in Fig.\ref{fig2a}, where the
evolutionary tracks of all the massive galaxies are shown: In group G1
and G2 all progenitors merge into the same object until redshift 0
whereas all but one galaxy in G3 merge to the massive central
galaxy. The timescales for those mergers with the central object are
mainly set by the initial distance (and orbit) from the central galaxy
and given the low mass ratios between the primary and the secondary
object ($M_s > 0.1 M_p$) the dynamical friction time-scale is bound to
be very short, even accounting for the effect of tidal mass loss,
($T_{DF}< 1$ Gyr) (Taffoni et al. 2003). Therefore the infalling
galaxies spiral in to the central in just one/two orbits after they
enter the main halo. The only companion galaxy surviving to redshift
zero does not merge because it forms far enough from the primary of
group G3 to enter the virial radius only shortly before redshift zero.
We note that those mergers are not completely dry despite the gas
removal by tidal and ram pressure stripping: the massive galaxies
merging to the central object are showing gas contents
$m_{gas}/m_{baryons}$ of $\sim 1 \%$ to $\sim 9 \%$ within the inner
10 kpc measured at the time when the distance between the main galaxy
and the infalling secondary falls below $30$ kpc for the first
time. When measured at the time when satellites enter for the first
time the virial radius of the main object, the gas fractions are $\sim
2 \%$ to $\sim 18\%$, with satellites crossing the virial radius at
high redshift $(z \sim 1.5)$ showing the highest gas fractions and
those entering late $(z < 0.8)$ having the lowest ones. The low gas
fractions might reflect the excessive star formation in the poorly
resolved centres of the galaxies (see also F10). An excessive star
formation is even more problematic in simulations adopting weak
feedback. For example, Oser et al. (2010) argue that the weak feedback
prescription used by Naab et al. (2007, 2009) artificially enhances
gas consumption in all galaxies at early times and speculate that the
inclusion of blastwave SN feedback such as ours would alleviate this
problem. The presence of significant gas components in some of the
satellites down to low redshift is a reassuring aspect of our
simulations, although, owing to the use of a low star formation
density threshold and a relatively low resolution in the gas phase
relative to recent zoom-in simulations of lower mass objects (Guedes
et al. 2011), the effect of feedback is likely still
underestimated. To take the effects of excessive star formation in the
poorly resolved centres of the galaxies into account, we show the
specific star formation rates (SSFR) $M_*^{-1}\,dM_*/dt $ measured
within a sphere of 10 kpc but excluding all the star formation
occurring within the inner sphere with radius of one softening length
(as in F10 and FCM11) in Fig. \ref{fig3}. In our sample of galaxies at
redshifts $2, 1.5, 0.7$ and $0$ star formation is generally more
efficient at higher redshift and for star forming galaxies ($SSFR >
10^{-2}$ Gyr$^{-1}$) fairly constant versus mass at a given redshift
(see e.g. Peng et al. 2010). The SSFR found at redshift 0 lie in the
lower part of observational findings for the local Universe (Salim et
al. 2007, see Fig. \ref{fig3}) and we note that at redshift 0 several
galaxies have a very low SSFR and are basically not star forming
anymore.

\begin{figure}
  \includegraphics[scale=1.0]{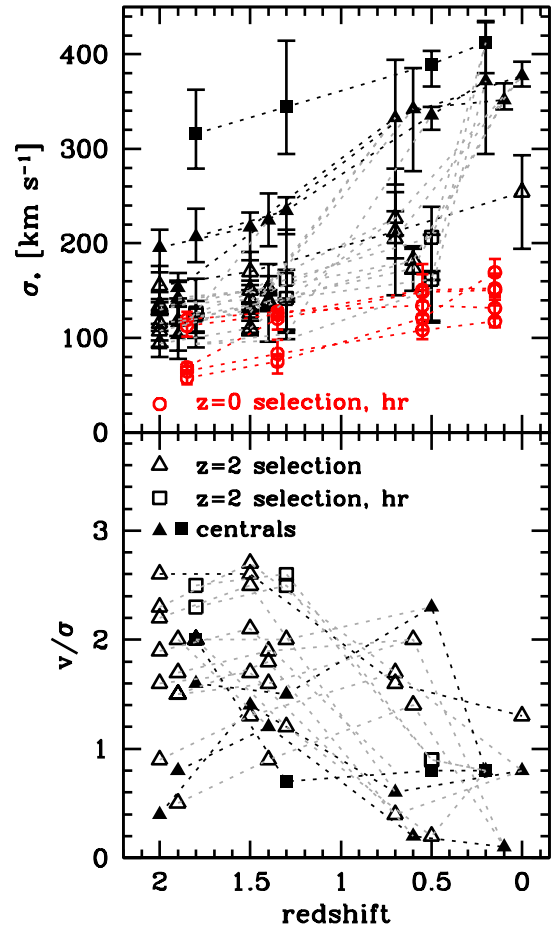}
  \caption{The time evolution of stellar velocity dispersion within
    the effective stellar radius of the massive galaxy population
    selected at redshift 2 (black) and at redshift 0 (red) is shown in
    the upper plot. In the bottom plot the rotational support of the
    massive galaxies selected at redshift 2 ($v/\sigma$ measured for
    stars within the effective radius) measured at redshift 2, 1.5,
    0.7 and 0 is shown. Measurements were taken at the redshifts
    indicated above but are plotted shifted slightly along the x-axis
    for better visibility. The measurement for the galaxies of the
    three different simulations are plotted using triangles (standard
    resolution) or squares (high resolution). Filled symbols are used
    for the galaxies evolving into the most massive, central object
    until redshift 0 in each of the simulations (the 'main
    progenitors'). Circles are used for the population of massive
    (satellite) galaxies selected at redshift 0 in the high resolution
    simulation. The high fraction of highly rotationally supported
    objects is decreasing with time.  \label{fig2b}}
\end{figure}

\begin{figure*}
\includegraphics[scale=0.7]{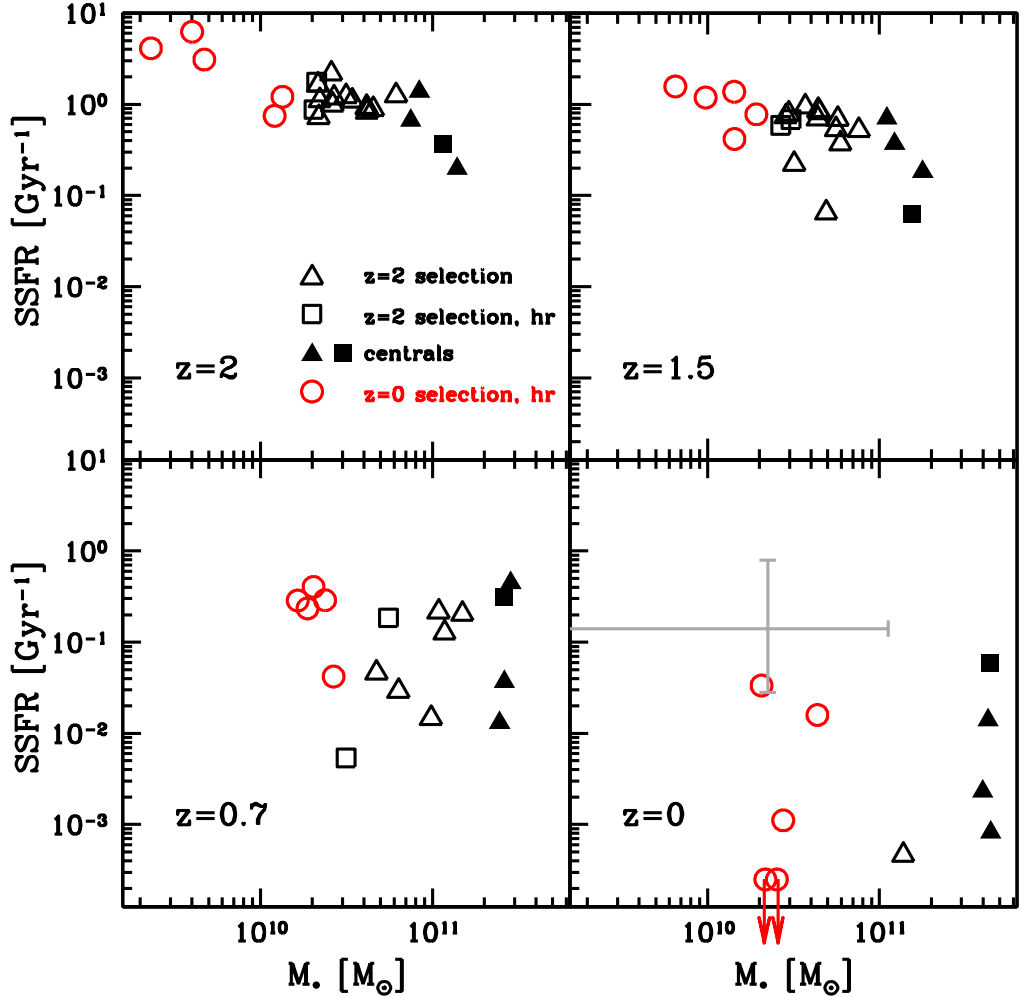}
\caption{The specific star formation rate (SSFR) $M_*^{-1}\,dM_*/dt $
  measured within a sphere of 10 kpc is plotted versus stellar mass
  $M_*$ for all the galaxies at redshifts $2, 1.5, 0.7$ and $0$. The
  inner sphere with radius one softening length has been excluded for
  the measurement of the SSFR as done in F10. The SSFR is decreasing
  with redshift. The gray error bars in the lower right plot indicate
  the contour encompassing $70\%$ of all the star-forming objects of
  Salim et al. (2007). \label{fig3}}
\end{figure*}

A high fraction of galaxies is highly rotationally supported
(disc-like) at redshift 2 (see Fig. \ref{fig2b}), similar as seen in
observations of massive galaxies at $z \sim 2$ (van der Wel et
al. 2011). The amount of rotational support decreases over time for the
simulated galaxies, likely due to mechanisms as mergers and tidal
stirring. Tidal stirring, namely repeated tidal shocks due to close
encounters with the central galaxy of the group (Mayer et al. 2001),
begins to operate after galaxies have entered the virial radius of the
main galaxy, see Fig. \ref{fig2b}. Tidal stirring can only become
effective after redshift 1.5, once a significant number of galaxies
have entered the virial radius of the respective main galaxy (see also
FCM11 for further environmental effects). While the fraction of highly
rotationally supported (disc-like) objects is decreasing as the
galaxies are falling into higher densities environments (which seems
to agree with the observed morphology-density relation, e.g. Postman
and Geller 1984, Goto et al. 2003) we caution that due to the
selection of the galaxies fixed at redshift 2 (and since we neglect
galaxies that cross the mass threshold at a later time) the derived
fractions of various classes of rotational support are not directly
comparable with observational mass-selected samples.

Fig. \ref{fig2a} and \ref{fig2b} show the evolution of stellar masses,
effective radii, stellar densities, stellar velocity dispersions and
rotational support ($v/\sigma$ measured for effective radius) at
redshift 2, 1.5, 0.7 and 0. Fig. \ref{fig2a} demonstrates that the
spread in mass in the progenitors of a factor $\sim 6$ at redshift 2
disappears and those galaxies evolve into a homogeneous (with respect
to mass) population of central galaxies at redshift 0. The
high-resolution run produces comparable masses as the standard
run. The one object of G3 which does not merge with the central object
grows in stellar mass as well but stays a factor of $\sim 2.5$ lower
in mass.

All the main progenitors show an effective radius $<0.7$ kpc
(Fig. \ref{fig2a}). Group G2 shows besides the massive central galaxy
only two galaxies above the mass-cut at $z =2$, both showing low
velocity dispersion and low stellar mass. The effective radii of the
central galaxies grow then significantly until redshift 0, mostly by
acquiring a stellar envelope (see F10 and also Szomoru et al. 2011).
While all three simulations end up with a central galaxy of similar
mass at redshift 0 the evolutionary paths were rather different as
shown in F10. The central galaxies of groups G1 and G3 both
experienced two major galaxy mergers between $z\sim 1.5$ and
$0$. Those major mergers add significant amounts of stellar mass to
the central galaxies, see Fig. 7 in F10. The central galaxy of group
G2 does not experience any major merger during that epoch and also not
below $z\sim 4$. It grew from minor mergers and in situ star formation
(see also Oser et al. 2011).  In Fig.  \ref{fig2b} the stellar
velocity dispersions are shown. At redshift 2 the range in velocity
dispersion covers a wide range from high stellar velocity dispersion
for compact massive galaxies to values typical for intermediate
ellipticals.

The $z>0$ galaxies are generally more compact (i.e. smaller effective
radii and higher stellar velocity dispersion versus a given mass) than
the local sample of quiescent galaxies in SDSS as shown by van de
Sande et al. 2011 and follow more closely the observational data of
(massive, compact) $z>1$ galaxies compiled by those authors (see also
Newman et al. 2010). For the progenitor of today's satellite
population in simulation G2, see next section.

\subsection{Tracing today's satellites backwards in time}
\label{sat_0}

FCM11 found the progenitors of the $z=0.1$ satellite population to be
discy, blue, gas-rich star forming galaxies, which have assembled half
their mass as late as $z\sim1.5$ to $z\sim 1$. We find additionally,
that the $z=0$ massive satellite population is less compact (has
higher effective radii in the same mass bin, smaller effective
densities and also have smaller maximal masses and smaller maximal
velocity dispersions, see Fig. \ref{fig1}) than the massive galaxies
selected at redshift 2. Tracing those satellites back in time shows
that at redshift 1.5 the stellar masses of all those satellites were
less $2 \times 10^{10}\Msun$ (the mass threshold needed for selection
at $z=2$) and two were less massive than $10^{10}\Msun$ (Fig.
\ref{fig2a}).  At redshift 0 four out of the five are residing within
the virial radius of the main object (one of these galaxies being at
it apocenter outside $R_{vir}$), whereas at $z=1.5$ none of them was
within the virial radius of the main galaxy.  At $z=2$ the progenitors
of those satellites were forming by mostly ``in situ'' star formation
(see also Oser et al. 2010) far away from the main galaxy at
distances ranging from $\sim 7$ to $\sim 14$ times the virial radius
(at $z=1.5$ from $\sim 4$ to $>10$ times the virial radius) of the
most massive galaxy, being farther away than the massive galaxies
selected at redshift $2$ and also farther away than the turnaround
radius at those times ($R_t \sim 3.5 \, R_{vir}$, see Cupani et
al. 2008). Therefore these galaxies fell in later\footnote{These
  galaxies seem to be part of a ``second generation'' of massive
  galaxies: forming later in the outskirts, thus reaching the central
  area of halo later.} than the massive galaxies selected at redshift
2.  Since the virial masses of the selected satellites are typically a
factor 20 to 100 smaller than the one of the central galaxy, dynamical
friction is then not strong enough to merge those satellites with the
central (Taffoni et al. 2003) once they fell into the virial radius of
the central and the satellites are predicted to settle into inner
orbits.  Note that at $z=2$, especially for the low mass objects, the
low number of stellar particles of those galaxies made the
measurements the stellar velocity dispersion and effective radii less
reliable than for the more massive galaxies. Also, some of these
galaxies just formed and are about to collapse further (i.e., to
smaller effective radii). These two effects become apparent especially
in Fig.  \ref{fig2a} where the large effective radii for low-mass
objects increase the size of the parameter space substantially.

\section{The role of formation time, numerical resolution and missing
  physics in shaping structural properties of galaxies}

\subsection{Formation time and compactness}
We note that the most peculiar object in our sample, the most massive
galaxy in the simulation G2 at $z=2$ (i.e. the most massive progenitor
of the central object at $z=0$) is the most massive galaxy in our
$z=2$ sample and also is the object in which star formation begins
earlier than in any other galaxy, at $z > 5.5$ (a statement
independent on resolution since it is true in both the standard and
high resolution simulation of G2). In groups G1 and G3 the objects
which formed stars first did not merge until after $z=2$ to the later
central object. Also, if the formation time of a galaxy is defined as
the time when the object has acquired $20\%$ of its stellar mass at
$z=0$ we find that massive galaxy in G2 forms the earliest, at $z\sim
2.1$, whereas the central objects of G1 and G3 form at $z\sim1.5$. The
choice of the 20\% criterion is somewhat heuristic but appears to
reflect conservatively the rate of stellar mass build up in the
different galaxies (smaller reference mass fractions to define
formation time highlight even more the correlation between stellar
mass at $z=2$ and formation time).  At the respective formation epoch
the stars of the main galaxies can also be characterised as sitting at
the deepest points of the potential well: a fraction of $0.80, 1.00 $
and $0.79 $ of the main galaxy stars are among the $20\%$ of the stars
having the lowest potential in the whole G1, G2 and G3 simulation,
respectively.

After redshift $\sim 1.5$ the mass in the central $2$ kpc of the
central galaxies stays roughly constant but their total mass increases
by a factor of $3-4$. The new stellar material is accreted (or formed in-situ)
outside the central region (F10). Therefore the effective radii
increase until $z=0$. Another main mechanism for increasing the
effective radius at work in the simulations are mergers, both minor and major, 
as described in Naab et al. (2009). Typically our massive satellites, after a
first decrease of the effective radius during the formation phase, 
undergo an increase of size owing to all the mechanisms just mentioned,
except that mergers play a more important role in the most massive progenitors 
of the central galaxy. Furthermore, since the satellites selected at $z=2$ merge with the central
galaxy, and the galaxies selected at $z=0$ do not have enough time to
grow until $z=0$ due to their late formation, the overall effect of the various
mechanisms behind size growth is weaker than in the centrals.

\subsection{Resolution tests and potential additional physics}
The results presented in the last section might be affected by missing
numerical resolution (our standard runs do not quite reach the number
of particles needed for convergence as suggested in Naab et al. (2007)
but the high-resolution run does). Stellar velocity dispersion can in
principle be inflated by two-body heating from massive halo particles,
but can also be lowered by a larger softening length (as it might be
the case when comparing G2 at standard and high resolution).  Also,
all cosmological simulations of galaxy formation are suffering from
e.g. artificial angular momentum loss since galaxy progenitors at
high-z are poorly resolved (Kaufmann et al. 2007); this could lead to
an artificially increased velocity dispersion and higher central
densities. The only cosmological simulation existing to date that
satisfies the resolution criteria of Kaufmann et al. (2007) is the
ERIS simulation (Guedes et al. 2011) which however follows the
formation of a galaxy in a halo almost 30 times less massive that
those considered here.  On the other hand the star formation and
feedback parameters have been kept fixed while increasing the
resolution, while recent results on smaller mass scales show that
central densities are influenced by the appropriate choice of these
parameters, in particular the star formation density threshold (see
e.g. Governato et al. 2010, Guedes et al. 2011). Comparing the results
of our standard G2 simulation with the high resolution G2 run we found
that the mass of the respective objects agree quite well
(Fig. \ref{fig2a}) but the velocity dispersion in the high resolution
run was found to be somewhat larger ($ + 24 \%$) and the effective
radius smaller ($ - 29 \%)$ than in the standard run measured at
$z=2$. This shows that, if anything, we are erring on the side of
underestimating the compactness of our galaxies, and suggests that we
are not dominated by numerical angular momentum loss but rather by
e..g. softening, that tends to reduce densities and typical central
velocities. Most importantly, since all these effects will be present
mostly at high z, we argue that the {\em trends} versus time
identified in this work are fairly robust.

Additional physical effects not implemented in our simulations, such
as feedback from active galactic nuclei (AGN), might play an important
role in the formation of massive galaxies, especially for more massive
systems than the ones studied here (Naab et al. 2007, Teyssier et
al. 2011). F10 argue against the a major role of AGN feedback at the
mass of groups because early-type galaxies at $z =0$ have global
properties close to those of observed galaxies, with only central
densities and total stellar masses somewhat on the high side relative
to typical observed early-types. A mild effect of AGN feedback, which
mostly self-regulates star formation at the centre but does not drive
strong baryonic outflows, might be enough to solve the problem. Such a
scenario might be the closest to reality, perhaps more realistic than
the popular "quasar mode" feedback, according to recent galaxy
simulations that implement directly radiative transfer of the X-ray
and ionising radiation released as a result of accretion onto the
central AGN (Kim et al. 2011).  A very high resolution cosmological
simulation with the inclusion of an educated model of AGN feedback,
along with more realistic modelling of the multi-phase interstellar
medium and subsequent star formation processes (Robertson \& Kravtsov
2008), will be needed to really assess the importance of AGN feedback
in the evolution of early-type galaxies.

\section{Summary and  Conclusions}

We analysed three "zoom-in" cosmological simulations where the main
haloes have final virial masses of $\sim 10^{13}\Msun$ and host a
massive central early-type galaxy at $z=0$ (F10). We identified 16
massive galaxies at high redshift in those three simulations, which
fall into the main halos sooner or later. We showed that at redshift
$\sim 2$ the population of massive galaxies is very diverse in terms
of mass, velocity dispersion and effective radius.  A high fraction of
them has significant rotation, is disk-like and gas-rich, lending
theoretical support to the observational results by van der Wel et
al. (2011), who argue that the majority of the compact massive
galaxies at $z\sim2$ are disc dominated based on the shapes inferred
from the photometry (see also Genzel et al. 2006; 2008).  In
particular, the main progenitors of the central galaxies at $z=0$ are
objects with a sizable disc component; all but one of the galaxies
have a mean $v/\sigma > 0.5$ at $z \sim 2$.

The population of galaxies at $z=2$ comprises relatively massive
members (from $\sim 2 \times 10^{10}\Msun$ to $1.15 \times
10^{11}\Msun$) living in over-dense, yet unbound associations of order
a Mpc in size that will later assemble the potential of a virialised
group (for detailed results on the central objects see F10).  While we
find one compact, massive object resembling the extreme galaxy
described in van Dokkum et al. (2009) no galaxy with low velocity
dispersion and large effective radius comparable to typical early type
galaxies at low redshift, or to the object described in Onodera et
al. (2010), has been identified in our sample at $z=2$. Despite the
diversity in properties, masses and merger histories of the galaxies
in the sample at $z=2$, the central galaxies at $z=0$, which are
assembled partly by merging and interactions between the $z=2$
progenitors and partly by accretion of gas and in situ star formation,
at $z=0$ end up roughly with the same stellar mass, a few $\times
10^{11}\Msun$, and with relatively similar morphologies (F10).

In all the three simulations only the most massive progenitor survives
until $z=0$, since the other massive galaxies identified at $z=2$
merge with it (with the exception of simulation G3 where one
additional galaxy survives).  The massive satellites selected at $z=0$
(less compact than their $z=2$ counterparts) were not physically
associated with the massive galaxies at $z=2$ since they became bound
to the growing group potential later on (their stellar mass also
crossed our threshold for selection much later).

Our findings suggest that is not surprising that today's massive
galaxy population is generally less compact than a population of
similar stellar mass at $z=2$; the compact population identified at
$z\sim2$ simply does not exist anymore today and no new galaxies form
with similar structural properties in the meantime.  They either
became the most massive, central galaxy (several mechanisms as
mergers, acquisition of a stellar envelope can increase the radii of
such galaxies) or they merged with the central object. Today's massive
satellite population formed later far from the central galaxy (at a
distance corresponding to $> 7$ times its virial radius at $z=2$) and
with relatively small masses that yield long dynamical friction times,
preventing merging with the central galaxy. Moreover, they form at a
time where typical characteristic densities at virialisation are much
lower than for galaxies already in place at $z > 2$, which naturally
explains their typical characteristic densities.

Also the observed spread in velocity dispersion of massive galaxies at
high redshift is easily understandable in a $\Lambda$CDM Universe
since galaxies at e.g. $z = 2$ might be at different points in
assembling a substantial fraction of their final mass. This is mostly
the result of the different assembly history of the central galaxies
at $z=0$, which is the criterion that we originally used in F10 to
select the candidate halos at a given mass scale at $z=0$ to
re-simulate at higher resolution with the zoom-in technique.

These findings are in agreement with the picture of a diverse
population of massive galaxies at $z>1$ put forward by the
observational work of van Dokkum et al. (2011). Similarly, the results
of our simulations are along the lines of the findings of Cassata et
al. (2011)(massive passively evolving early-type galaxies form compact
for redshifts $>1$, growing later in size, and late forming early-type
galaxies are larger\footnote{Note, that the majority of our massive
  galaxies are still forming stars at high redshift.}). We also note
that our results on the size growth of the central galaxies ($M_*\sim
10^{11}\Msun$ at $z=2$) agree with the observational findings of
Carollo et al. (in preparation) and also with the simulation work of
Oser et al. (2011). However, we argue that our work discusses massive
galaxy evolution in a more typical setting than the latter, since by
choosing a group environment rather than field objects we have been
able to study the evolution of massive galaxies that are not centrals,
either at high or low redshifts, an evolution that is driven by
environmental effects.  For a population of field galaxies we do not
expect a similar growth in the average size of the population to occur
(due to destruction of the compact population by merging) simply
because the average merger rate should be lower for field objects.
Such an environmental dependence of the mass-size relation has already
been suggested by observational work (Raichoor et al. 2011, Cooper et
al. 2011). Therefore, our work and that of Oser et al. (2011) can be
viewed as complimentary.

While our results do not reflect a statistical analysis of thousands
of galaxies we argue that they should be fairly general for relatively
dense environments. Indeed the three simulations were chosen to have
different merger histories and larger scale environments, ranging from
G1 being an isolated group in the cosmic web to G3 having a nearby
cluster and two other virialised groups within 5 Mpc (F10). Most
importantly, haloes at the $10^{13} \Msun$ mass scale are fairly
representative of dense environments at low redshift, namely the
environments in which massive galaxies are more common (see Eke et
al. 2004).  The number density for haloes with masses $5\times
10^{12}$ to $2\times 10^{13} h^{-1} \Msun $ is $\sim 8 \times
10^{-4}\,h^3\,$Mpc$^{-3}$ based on the works of Macci{\`o} et
al. (2007). For comparison, the number density for all cluster-sized
haloes with $M>10^{14} h^{-1} \Msun$ is only $\sim 3 \times
10^{-5}\,h^3\,$Mpc$^{-3}$ (and $N(>2\times 10^9 h^{-1} \Msun) \sim 1
\,h^3\,$Mpc$^{-3}$).

All the ultra massive galaxies ($M_*>10^{11}\Msun$) in our sample, as
well as in the Oser et al. (2011) sample, were compact at $z=2$, and
no object at $z=2$ has been observed in those two samples with
properties similar to local giant ellipticals, like the object
described in Onodera et al. (2010). We argue that we miss such a
galaxy in our sample because of the selection imposed by our halo mass
scale at $z=0$.  We speculate that the ultra massive extended galaxies
might be born earlier in halos of even higher masses. They would
therefore begin to grow in mass and size earlier and develop lower
concentrations by $z=2$ (the galaxy in Onodera et al. 2010) has
indeed has a stellar mass higher than that of our sample at $z=2$).
Such galaxies should end up in the cluster potentials at $z=0$. A
future test of this idea would be a comparison of the number density
of, respectively, dense giant ellipticals and extended giant
ellipticals at redshift $\sim 2$, with the number densities of,
respectively, group and cluster haloes in a $\Lambda$CDM universe.

\section*{Acknowledgements}

T. K. acknowledges financial support from the Swiss National Science
Foundation (SNF).




\begin{thebibliography}{}

\bibitem[Bertschinger(1995)]{1995astro.ph..6070B} Bertschinger, E.\ 1995, 
arXiv:astro-ph/9506070 

\bibitem[Bertschinger(2001)]{2001ApJS..137....1B} Bertschinger, E.\ 2001, 
\apjs, 137, 1 

\bibitem[Bezanson et al.(2009)]{2009ApJ...697.1290B} Bezanson, R., van 
Dokkum, P.~G., Tal, T., Marchesini, D., Kriek, M., Franx, M., 
\& Coppi, P.\ 2009, \apj, 697, 1290 

\bibitem[Bluck et al.(2011)]{2011arXiv1111.5662B} Bluck, A.~F.~L., 
Conselice, C.~J., Buitrago, F., et al.\ 2011, arXiv:1111.5662 

\bibitem[Brooks(2010)]{2010ASPC..432...17B} Brooks, A.\ 2010, New
  Horizons in Astronomy: Frank N.~Bash Symposium 2009, 432, 17

\bibitem[Cappellari et al.(2009)]{2009ApJ...704L..34C} Cappellari, M.,
  et al.\ 2009, \apjl, 704, L34

\bibitem[Cassata et al.(2011)]{2011arXiv1106.4308C} Cassata, P., et al.\ 
2011, arXiv:1106.4308 

\bibitem[Cooper et al.(2011)]{2011MNRAS.tmp.1893C} Cooper, M.~C.,
  Griffith,  R.~L., Newman, J.~A., et al.\ 2011, \mnras, 1893 

\bibitem[Cupani et al.(2008)]{2008MNRAS.390..645C} Cupani, G., Mezzetti, 
M., \& Mardirossian, F.\ 2008, \mnras, 390, 645 

\bibitem[Daddi et al.(2005)]{2005ApJ...626..680D} Daddi, E., et al.\ 2005, 
\apj, 626, 680 

\bibitem[Eke et al.(2004)]{2004MNRAS.348..866E} Eke, V.~R., Baugh, C.~M., 
Cole, S., et al.\ 2004, \mnras, 348, 866 

\bibitem[Feldmann et al.(2010)]{2010ApJ...709..218F} Feldmann, R., Carollo, 
C.~M., Mayer, L., Renzini, A., Lake, G., Quinn, T., Stinson, G.~S., 
\& Yepes, G.\ 2010, \apj, 709, 218 

\bibitem[Feldmann et al.(2011)]{2011ApJ...736...88F} Feldmann, R., Carollo, 
C.~M., \& Mayer, L.\ 2011, \apj, 736, 88 

\bibitem[Genzel et al.(2006)]{2006Natur.442..786G} Genzel, R., et al.\ 
2006, \nat, 442, 786 

\bibitem[Genzel et al.(2008)]{2008ApJ...687...59G} Genzel, R., et al.\ 
2008, \apj, 687, 59 

\bibitem[Governato et al.(2010)]{2010Natur.463..203G} Governato, F., et 
al.\ 2010, \nat, 463, 203 

\bibitem[Goto et al.(2003)]{2003MNRAS.346..601G} Goto, T., Yamauchi, C., 
Fujita, Y., Okamura, S., Sekiguchi, M., Smail, I., Bernardi, M., 
\& Gomez, P.~L.\ 2003, \mnras, 346, 601 

\bibitem[Guedes et al.(2011)]{2011arXiv1103.6030G} Guedes, J., Callegari, 
S., Madau, P., \& Mayer, L.\ 2011, arXiv:1103.6030 

\bibitem[Hahn et al.(2007)]{2007MNRAS.375..489H} Hahn, O., Porciani, C., 
Carollo, C.~M., \& Dekel, A.\ 2007, \mnras, 375, 489 

\bibitem[Hopkins et al.(2010)]{2010MNRAS.401.1099H} Hopkins, P.~F., Bundy, 
K., Hernquist, L., Wuyts, S., \& Cox, T.~J.\ 2010, \mnras, 401, 1099 

\bibitem[Kaufmann et al.(2007)]{2007MNRAS.375...53K} Kaufmann, T.,
  Mayer, L., Wadsley, J., Stadel, J., \& Moore, B.\ 2007, \mnras, 375,
  53

\bibitem[Kennicutt(1998)]{1998ApJ...498..541K} Kennicutt, R.~C., Jr.,
  1998, \apj, 498, 541

\bibitem[Khochfar \& Silk(2006)]{2006ApJ...648L..21K} Khochfar, S., \&
  Silk, J.\ 2006, \apjl, 648, L21

\bibitem[Kim et al.(2011)]{2011ApJ...738...54K} Kim, J.-h., Wise, J.~H., 
Alvarez, M.~A., \& Abel, T.\ 2011, \apj, 738, 54 

\bibitem[Macci{\`o} et al.(2007)]{2007MNRAS.378...55M} Macci{\`o},
  A.~V., Dutton, A.~A., van den Bosch, F.~C., Moore, B., Potter, D.,
  \& Stadel, J.\ 2007, \mnras, 378, 55

\bibitem[Mancini et al.(2010)]{2010MNRAS.401..933M} Mancini, C., et al.\ 
2010, \mnras, 401, 933 

\bibitem[Mayer et al.(2001)]{2001ApJ...547L.123M} Mayer, L., Governato, F., 
Colpi, M., Moore, B., Quinn, T., Wadsley, J., Stadel, J., 
\& Lake, G.\ 2001, \apjl, 547, L123 

\bibitem[Mayer et al.(2008)]{2008ASL.....1....7M} Mayer, L., Governato, F., 
\& Kaufmann, T.\ 2008, Advanced Science Letters, 1, 7 

\bibitem[M{\'a}rmol-Queralt{\'o} et al.(2012)]{2012arXiv1201.2414M} 
M{\'a}rmol-Queralt{\'o}, E., Trujillo, I., P{\'e}rez-Gonz{\'a}lez, P.~G., 
Varela, J., \& Barro, G.\ 2012, arXiv:1201.2414 

\bibitem[McKee \& Ostriker(1977)]{1977ApJ...218..148M} McKee, C.~F.,
  \& Ostriker, J.~P.\ 1977, \apj, 218, 148

\bibitem[Miller \& Scalo(1979)]{1979ApJS...41..513M} Miller, G.~E., \&
  Scalo, J.~M.\ 1979, \apjs, 41, 513

\bibitem[Naab et al.(2007)]{2007ApJ...658..710N} Naab, T., Johansson, 
P.~H., Ostriker, J.~P., \& Efstathiou, G.\ 2007, \apj, 658, 710 

\bibitem[Naab et al.(2009)]{2009ApJ...699L.178N} Naab, T., Johansson, 
P.~H., \& Ostriker, J.~P.\ 2009, \apjl, 699, L178 

\bibitem[Newman et al.(2010)]{2010ApJ...717L.103N} Newman, A.~B., Ellis, 
R.~S., Treu, T., \& Bundy, K.\ 2010, \apjl, 717, L103 

\bibitem[Newman et al.(2011)]{2011arXiv1110.1637N} Newman, A.~B., Ellis, 
R.~S., Bundy, K., \& Treu, T.\ 2011, arXiv:1110.1637 

\bibitem[Onodera et al.(2010)]{2010ApJ...715L...6O} Onodera, M., Daddi, E., 
Gobat, R., et al.\ 2010, \apjl, 715, L6 

\bibitem[Oser et al.(2010)]{2010ApJ...725.2312O} Oser, L., Ostriker, J.~P., 
Naab, T., Johansson, P.~H., \& Burkert, A.\ 2010, \apj, 725, 2312 

\bibitem[Oser et al.(2011)]{2011arXiv1106.5490O} Oser, L., Naab, T., 
Ostriker, J.~P., \& Johansson, P.~H.\ 2011, arXiv:1106.5490 

\bibitem[Peng et al.(2010)]{2010ApJ...721..193P} Peng, Y.-j., et al.\ 2010, 
\apj, 721, 193 

\bibitem[Postman \& Geller(1984)]{1984ApJ...281...95P} Postman, M., \&
  Geller, M.~J.\ 1984, \apj, 281, 95

\bibitem[Raichoor et al.(2011)]{2011arXiv1109.0284R} Raichoor, A.,
  Mei, S., Stanford, S.~A., et al.\ 2011, arXiv:1109.0284

\bibitem[Robertson \& Kravtsov(2008)]{2008ApJ...680.1083R} Robertson,
  B.~E., \& Kravtsov, A.~V.\ 2008, \apj, 680, 1083

\bibitem[Salim et al.(2007)]{2007ApJS..173..267S} Salim, S., et al.\ 2007, 
\apjs, 173, 267 

\bibitem[Spergel et al.(2007)]{2007ApJS..170..377S} Spergel, D.~N., et al.\ 
2007, \apjs, 170, 377 

\bibitem[\protect\citeauthoryear{Stadel} {2001}] {Stadel}Stadel J.,
  2001, PhD Thesis, U. Washington

\bibitem[Stinson et al.(2006)]{2006MNRAS.373.1074S} Stinson, G., Seth, A., 
Katz, N., Wadsley, J., Governato, F., \& Quinn, T.\ 2006, \mnras, 373, 1074 

\bibitem[Szomoru et al.(2010)]{2010ApJ...714L.244S} Szomoru, D., et al.\ 
2010, \apjl, 714, L244 

\bibitem[Szomoru et al.(2011)]{2011arXiv1111.3361S} Szomoru, D., Franx, M., 
\& van Dokkum, P.~G.\ 2011, arXiv:1111.3361 

\bibitem[Taffoni et al.(2003)]{2003MNRAS.341..434T} Taffoni, G., Mayer, L., 
Colpi, M., \& Governato, F.\ 2003, \mnras, 341, 434 

\bibitem[Teyssier et al.(2011)]{2011MNRAS.414..195T} Teyssier, R., Moore, 
B., Martizzi, D., Dubois, Y., \& Mayer, L.\ 2011, \mnras, 414, 195 

\bibitem[van der Wel et al.(2011)]{2011ApJ...730...38V} van der Wel, A., et 
al.\ 2011, \apj, 730, 38 

\bibitem[van de Sande et al.(2011)]{2011ApJ...736L...9V} van de Sande, J., 
Kriek, M., Franx, M., et al.\ 2011, \apjl, 736, L9 

\bibitem[van Dokkum et al.(2009)]{2009Natur.460..717V} van Dokkum, P.~G., 
Kriek, M., \& Franx, M.\ 2009, \nat, 460, 717 

\bibitem[van Dokkum et al.(2011)]{2011arXiv1108.6060V} van Dokkum, P.~G., 
Brammer, G., Fumagalli, M., et al.\ 2011, arXiv:1108.6060 

\bibitem[\protect\citeauthoryear{Wadsley et al.} {2004}]{Wadsley}
  Wadsley J., Stadel J., Quinn T., 2004, NewA, 9, 137

\bibitem[Whitaker et al.(2011)]{2011arXiv1112.0313W} Whitaker, K.~E., 
Kriek, M., van Dokkum, P.~G., et al.\ 2011, arXiv:1112.0313 

\bibitem[York et al.(2000)]{2000AJ....120.1579Y} York, D.~G., et al.\ 2000, 
\aj, 120, 1579 


\end{thebibliography}
\end{document}